\documentstyle[pre,aps,epsf,multicol]{revtex}
\begin{document}
\title{Diffusion Limited Aggregation and Iterated Conformal Maps}
\author{Benny Davidovitch,$^1$ H.G.E Hentschel,$^2$ Zeev Olami,$^1$ Itamar
Procaccia,$^1$
Leonard M. Sander$^3$ and Ellak Somfai$^3$} \address{$^1$ Dept.
of Chemical Physics, The Weizmann Institute of Science, Rehovot 76100, Israel\\
$^2$ Department of Physics, Emory University, Atlanta, GA, 30322\\
$^3$ H. M. Randall Laboratory of Physics, The University of Michigan, Ann
Arbor, MI, 48109}

\date{version of \today}
\maketitle

\begin{abstract}
The creation of fractal clusters by diffusion limited aggregation (DLA) is
studied by using iterated stochastic conformal maps following the method
proposed
recently by Hastings and Levitov. The object of interest is the function
$\Phi^{(n)}$ which conformally maps the exterior of the unit circle to the
exterior
of an $n$-particle DLA. The map $\Phi^{(n)}$ is obtained from $n$
stochastic iterations of
a function $\phi$ that maps the unit circle to the unit circle with a bump. The
scaling properties usually studied in the literature on DLA appear in a new
light
using this language. The dimension of the cluster is determined by the linear
coefficient in the Laurent expansion of $\Phi^{(n)}$, which asymptotically
becomes
a deterministic function of $n$. We find new relationships
between the generalized dimensions of the harmonic measure and the scaling
behavior
of the Laurent coefficients.
\end{abstract}
\pacs{PACS numbers: ...}
\begin{multicols}{2}
\section{Introduction}

The diffusion limited aggregation (DLA) model was introduced in 1981 by
T. Witten and L. Sander \cite{81WS}. The model has been shown to underlie many
pattern forming processes including dielectric breakdown \cite{84NPW}, two-fluid
flow \cite{84Pat}, and electro-chemical deposition \cite{89BMT}. The model
begins
with fixing one particle at
the center of coordinates in $d$-dimensions, and follows the creation of a
cluster by
releasing random walkers from infinity, allowing them to walk around until
they hit
any particle belonging to the cluster. Upon hitting they are attached to the
growing cluster. The model was studied on- and off- lattice in several
dimensions
$d\ge2$; here we are only interested in the off-lattice versions in two
dimensions.

DLA has attracted enormous interest over the years since it is a remarkable
example
of the spontaneous creation of fractal objects. It is believed that
asymptotically
(when the number of particles $n\to \infty$) the dimension $D$ of the cluster is
very close to 1.71 \cite{88Mea}, although there exists to date no accepted
proof
for this fact in spite of several interesting attempts \cite{95Piet,94Hals}.
In addition, the model has attracted interest since it was among the first
\cite{86HMP}
to offer a true multifractal
measure: the harmonic measure (which determines the probability that a random
walker from infinity will hit a point at the boundary) exhibits singularities
that are usefully described using the multifractal formalism \cite{86HJKPS}.
Nevertheless DLA still poses more unsolved problems than answers. It is obvious
that a new language is needed in order to allow fresh attempts to explain the
growth patterns, the fractal dimension, and the multifractal properties of the
harmonic measure.

Such a new language was proposed recently by Hastings and Levitov
\cite{98HL}, \cite{97Hast}. These
authors showed that DLA in two dimensions can be grown by iterating
stochastic conformal
maps. We adopt their basic strategy and will see that it provides a new
formulation of the problem which may lead to new insights and results.

The basic idea is to follow the evolution of the conformal mapping
$\Phi^{(n)}(w)$
which maps the exterior of the unit circle in the mathematical $w$--plane onto
the complement of the cluster of $n$ particles in the physical $z$--plane.
$\Phi^{(n)}$ is unique by the Riemann mapping theorem, provided that it satisfies
the boundary condition \begin{equation}
\Phi^{(n)}(w) \sim F^{(n)}_{1} w \quad{\rm as}\quad w \to \infty\,.
\label{eq-boundary}
\end{equation}
Here $F^{(n)}_{1}$ is a real positive coefficient, fixing the argument of
$[\Phi^{(n)}(w)]'$ to be zero at infinity. $\Phi^{(n)}(w)$ is related to
the complex electric potential $\Psi^{(n)}(z)$ by
\begin{equation} \Psi^{(n)}(z) = \ln {h^{(n)}} (z)\,,
\label{eq-potential}
\end{equation}
where $h^{(n)}(z) = [\Phi^{(n)}]^{-1} (z)$ is the inverse mapping.
Letting $z \rightarrow \infty$ in Eq.(\ref{eq-boundary}) it is easy to
verify that Eq.(\ref{eq-potential}) implies
\begin{equation}
\Psi^{(n)}(z) \sim \ln z \qquad\hbox{when}\qquad z\to\infty
\label{eq-potential-asym}
\end{equation}
as it should be at $d=2$.

The equation of motion for $\Phi^{(n)}(w)$ is determined recursively. The
choice of the initial map $\Phi^{(0)}(w)$ is rather flexible, and in this
paper we select (arbitrarily) an initial condition $\Phi^{(0)}(w) = w$. We
expect the asymptotic cluster to be independent of this choice. Then suppose
that $\Phi^{(n-1)}(w)$ is given. The cluster of $n$ ``particles'' is created
by adding a new ``particle'' of constant shape and linear scale
$\sqrt{\lambda_0}$ t
o the cluster of $(n-1)$ ``particles'' at a position
which is chosen randomly according to the harmonic measure. We denote points
on the boundary of the cluster by $Z(s)$ where $s$ is an arc-length
parametrization.
The probability to add a particle on an infinitesimal arc $ds$ centered
at the point $z(s)$ on the cluster boundary is \begin{equation}
P(s,ds) \sim |\nabla \Psi (s)|ds \,. \label{psds} \end{equation} The pre-image
of $z(s)$ and $ds$ in the $w$-plane are $e^{i \theta}$ and $d\theta$
respectively.
Clearly, $ds =
|[\Phi^{(n-1)}]'(e^{i\theta})| d\theta$. From Eq.(\ref{eq-potential}) we
conclude that
\begin{equation}
P(s,ds) \sim | \nabla \Psi (s) ||\Phi'| d\theta = d\theta \,, \label{dteta}
\end{equation}
so the harmonic measure on the real
cluster translates to a uniform measure on the unit circle in the
mathematical plane.

The image of the cluster of $n$ particles under $h^{(n)}(z)$ is, by definition,
just the unit circle. On the other hand, the image of the cluster of $n$
particles
under $h^{(n-1)}(z)$ is the unit circle with a small bump whose linear scale
is $\sqrt{\lambda_0} / |{\Phi'}^{(n-1)} (e^{i \theta_n})|$ where $e^{i
\theta_n}$
is the image (under $h^{(n-1)}$) of the point $z_n$ on the real cluster at
which the growth occurred.

\begin{figure}
\epsfxsize=8.6cm
\centerline{
\epsfbox{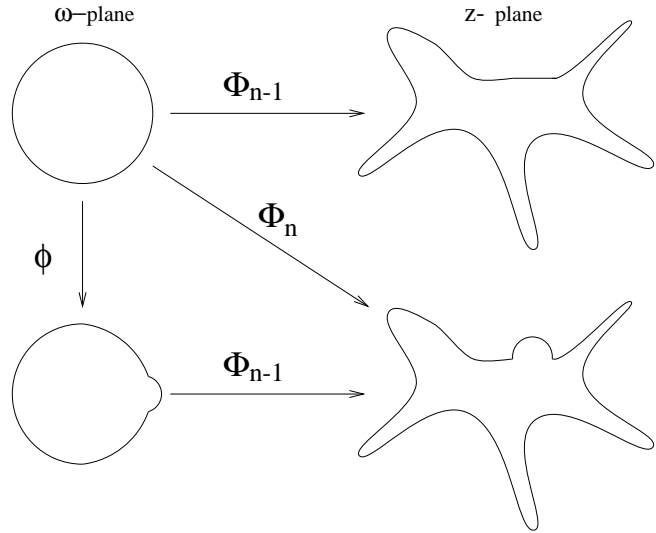}}
\caption{Diagramatic representation of the mappings $\Phi$ and $\phi$. }
\label{fig-mapping}
\end{figure}

Let us define now a new function $\phi_{\lambda_{n}, \theta_{n}}(w)$. This
function maps the unit circle to the unit circle with a bump of linear scale
$\sqrt{\lambda_{n}}$ around the point $e^{i \theta_n}$. For $w \rightarrow
\infty$,
$\phi_{\lambda_{n},\theta_{n}}(w) \sim w$ (with positive real
 proportionality coefficient). Using $\phi_{\lambda_{n},\theta_{n}}(w)$
the recursion relation for $\Phi^{(n)}(w)$ is given by (see
Fig.~\ref{fig-mapping}):
\begin{equation}
\Phi^{(n)}(w) = \Phi^{(n-1)}(\phi_{\lambda_{n},\theta_{n}}(w)) \ .
\label{eq-recursion}
\end{equation}

According to the above discussion $\lambda_{n}$ is given by \begin{equation}
\lambda_{n} = \frac{\lambda_0}{|{\Phi^{(n-1)}}' (e^{i \theta_n})|^2}
\label{eq-lambda}
\end{equation}
so the RHS of Eq.(\ref{eq-recursion}) is determined completely by
$\Phi^{(n-1)}(w)$; Eq.(\ref{eq-recursion}) induces the recursive dynamics
of $\Phi^{(n)}(w)$.

The recursive dynamics can be represented as iterations of the map
$\phi_{\lambda_{n},\theta_{n}}(w)$,
\begin{equation}
\Phi^{(n)}(w) =
\phi_{\lambda_1,\theta_{1}}\circ\phi_{\lambda_2,\theta_{2}}\circ\dots\circ
\phi_{\lambda_n,\theta_{n}}(\omega)\ . \label{comp} \end{equation} This
 composition appears as a standard iteration of stochastic maps. This is not so.
The order of iterations is inverted -- the last point of the trajectory is
the inner argument in this iteration. As a result the transition from
$\Phi^{(n)}(w)$ to
$\Phi^{(n+1)}(w)$ is not achieved by one additional iteration, but by
composing the
$n$ former maps Eq.(\ref{comp}) starting from a different seed which is no longer
$\omega$ but $\phi_{\lambda_{n+1},\theta_{n+1}}(w)$.

We note that in the physical plane the ``particles'' are roughly of the
same size.
To achieve this the linear scales $\sqrt{\lambda_n}$ vary widely as a
function of $n$ and $\theta$. We will see that the distribution of
$\sqrt{\lambda_n}$ and their correlations
for different values of $n$ determine many of the scaling properties of the
resulting cluster. In particular their moments are related to the generalized
dimensions of the harmonic measure.

There are many functions $\phi_{\lambda,\theta}$ which conformally map the
unit circle to the unit circle with a bump. A simple choice is a function which
behaves linearly for large $w$ and has a simple pole inside the unit circle
which
will induce a bump in the image. The pole has to be at $w_0=1-\lambda$ in order
to localize the bump near $w=1$ and make it of linear size of the order
$\sqrt{\lambda}$. The residue has to be $\lambda^{3/2}$,
in order for the bump's height to be also of the order $\sqrt{\lambda}$.
Consider
then $$\phi(w) = (1+ \lambda)w
+ \frac{\lambda^{3/2}}{w - w_0} \,.$$
Careful thinking leads to the conclusion that this function and other similar
functions are inappropriate: they have long ``tails''. In other words, the unit
circle is slightly distorted everywhere. This small global distortion may
result
in a loss of conformality or in the growth of non-constant size particles
in the
physical plane in numerical applications.

It was proposed in Ref.~\cite{98HL} that a choice for
$\phi_{\lambda_{n},\theta_{n}}(w)$
that is free of global distortion is given by
\begin{eqnarray}
&&\phi_{\lambda,0}(w) = w^{1-a} \left\{ \frac{(1+ \lambda)}{2w}(1+w)\right.
\nonumber\\
&&\left.\times \left [ 1+w+w \left( 1+\frac{1}{w^2} -\frac{2}{w} \frac{1-
\lambda}
{1+ \lambda} \right) ^{1/2} \right] -1 \right \} ^a \\
&&\phi_{\lambda,\theta} (w)
= e^{i \theta} \phi_{\lambda,0}(e^{-i \theta} w) \,, \label{eq-f}
\end{eqnarray}
The parameter $a$ is confined in the range $0<a<1$. As $a$ decreases the
bump becomes
flatter, with the identity map obtained for $a=0$. As $a$ increases towards
unity
the bump becomes elongated normally to the unit circle, with a limit of
becoming
a line (``strike'' in the language of \cite{98HL}) when $a=1$. Naively one
might
think that the shape of the individual particle is irrelevant for the large
scale fractal statistics; we will see that this is not the case. The dependence
on $a$ is important and needs to be taken into account. Notice that this
map has
two branch points on the unit circle. The advantage of this is that the bump
is strongly localized. On the other hand repeated iterations of this map leads
to rather complex analytic structure.

The aim of this paper is therefore to investigate the scaling and statistical
properties of such iterated stochastic conformal maps with a view to discovering
the scaling properties induced by the dynamics which any analytic theory must
ultimately explain. In Section II we present the numerical procedure used to
generate the fractal clusters, and in Section III give the necessary
mathematical
background to describe such mappings. In particular we discuss the Laurent
expansion of the conformal map from the unit circle to the $n$-particle cluster;
the coefficients of the Laurent series have interesting scaling behavior with
the size of the cluster which is intimately related to the fractal dimension
of the cluster and to the generalized dimensions of the harmonic measure.
In Section IV we present
numerical results regarding the scaling properties of averages of the Laurent
coefficients and of the size parameter $\lambda_n$. The results are accompanied
by a theoretical
analysis and interpretation. In Section V we conclude with some remarks on
the road ahead.

\section{Numerical procedure}

The algorithm simulating the growth of the cluster is based on Ref. \cite{98HL}.
The $n$ ``particle'' cluster is encoded by the series of pairs
$\{(\theta_i,\lambda_i)\}_{i=1}^n$. Having the first $n-1$ pairs,
the $n^{\rm th}$ pair is found as follows: choose $\theta_n$ from a
uniform distribution in $[0, 2\pi]$, independent of previous history.
Then compute $\lambda_n$
from Eq.(\ref{eq-lambda}), where the derivative of the iterated function
$\Phi^{(n-1)}$ involves $\phi_{\lambda_{n-1},\theta_{n-1}}'$,
$\phi_{\lambda_{n-2},\theta_{n-2}}'$, $\phi_{\lambda_{n-3},\theta_{n-3}}'$
etc, computed respectively at the points $e^{i\theta_n},\, \phi_{\lambda_{n-1},
\theta_{n-1}}(e^{i\theta_n}),\, \phi_{\lambda_{n-2},\theta_{n-2}}
(\phi_{\lambda_{n-1},\theta_{n-1}} (e^{i\theta_n}))$, etc. Notice that
the evaluation of both $\phi'$ and $\phi$ after the addition of one
particle involves $O(n)$ operations since the seed changes at every $n$.
This translates into
$n^2$ time complexity for the growth of an $n$-particle cluster. This
is inferior to
the best algorithms to grow DLA
(using hierarchical maps \cite{Ball}, with close to linear efficiency), but
the present algorithm is not aimed at efficiency. Rather, it is used since
the Laplacian field and the growth probability which is derived from it are
readily available at every point of the cluster and away from it. The typical
time to grow a 10,000 particle cluster is 8 minutes on a 300 MHz Pentium-II.

\begin{figure}
\epsfysize=5 cm
\epsfxsize=15 cm
\centerline{
\epsfbox{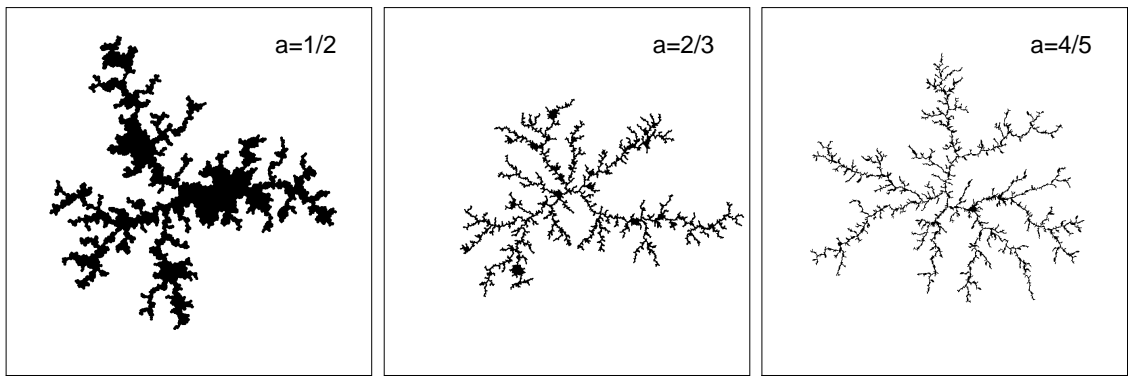}}
\bigskip
\caption{Typical clusters of 10,000 particles.
The black regions represent the interiors of the imagesof the unit circle
under the map $\Phi^{(10,000)}$ for three values of $a$.
Note the large enclosed area on the left branch.
}
\label{fig-clusters}
\end{figure}

Naively one would expect that any choice of $0<a<1$ would yield DLA clusters,
since $a$ only determines the shape of the particles (the aspect ratio is
$\frac{1}{2}a/(1-a)$ for small $\lambda$), and the microscopic details of the
particles (except their linear size) should not affect the global properties.
Three typical clusters with particles of various aspect ratios $a$ are shown
in Fig.~\ref{fig-clusters}. We mark in black the interior of the image of the
unit circle under
the conformal map $\Phi^{(n)}(w)$. The objects look very much like typical DLA
clusters grown by standard off-lattice techniques, and in the next section we
demonstrate that they have fractal dimensions in close agreement
with the latter. For $a$ significantly different from $2/3$, disadvantages of
the algorithm get amplified. Since the functional form of $\phi$ is fixed (only
the size and position of the ``bump'' change), particles of constant shape and
size are obtained only if the magnification factor $|{\Phi^{(n-1)}}'|$ (the
inverse of the field) is approximately constant in the $w$--plane around the
``bump'' of $\phi$. If the particles are elongated along the cluster, then the
variation of the field along the cluster affects the shape: large otherwise
deeply invaginated regions, where $\Phi'$ is large, are filled up with a single
particle, and the resulting cluster tends to be more compact. This effect,
slightly noticeable even at $a=2/3$, is quite significant at the otherwise
natural choice of $a=1/2$, where the particles are half circles. In
Fig.~\ref{fig-clusters}
we show such a cluster and point out to the area filling dark regions which
represent such unwanted
events. The other extreme, when the particles stick out of the cluster,
leads to
sensitivity to variations in the field going {\em away} from the cluster.
Especially if a bump is grown on a tip of a branch, where the field decreases
rapidly as one goes away from the tip (such that $\Phi'$ increases
significantly),
then the map of the bump gets magnified, resulting in particles of very
unequal sizes.

It is necessary to stress that even for $a=2/3$, when this procedure appears to
yield nice ramified structures, the problem of fill-ups does not go away:
in a few
rare cases the particle -- if it happens to land on a place where $|\Phi''|$
is large -- is significantly distorted. The net effect is that large areas
surrounded by the cluster (where the growth probability is small) are filled up
entirely by one distorted particle. For the value of $a=1/2$ it appeared that
the errors may be unbounded. Our numerics indicates that for $a=2/3$ the errors
were bounded for the cluster sizes that we considered. We do not have a
mathematical
proof of boundedness of the errors, and our disregard of this danger is only
based on the sensible appearance of our clusters at this value of $a$.
\section{Mathematical Background}

In this section we discuss the Laurent expansion of our conformal maps, and
introduce the statistical objects that are studied numerically in the next
section.

\subsection{Laurent Expansion}

Since the functions $\Phi^{(n)}(w)$ and $\phi_{\lambda,\theta}(w)$ are required
to be linear in $w$ at infinity, they can be expanded in a Laurent series in
which the highest power is $w$: \begin{eqnarray}
\Phi^{(n)}(w) &&= F^{(n)}_{1} w + F^{(n)}_{0} + F^{(n)}_{-1}w^{-1} +
F^{(n)}_{-2}w^{-2}
+ \dots
\label{eq-laurent-F} \\
\phi_{\lambda,\theta}(w) &&= f_1w+f_0 +f_{-1}w^{-1} + f_{-2}w^{-2}+ \dots
\label{eq-laurent-f}
\end{eqnarray}
where
\begin{eqnarray}
f_1 & = & (1+ \lambda)^a \nonumber\\
f_0 &=&\frac{2a\lambda \, e^{i \theta}}{(1+ \lambda)^{1-a}} \nonumber\\ f_{-1}
&=&\frac{2a\lambda \, e^{2i\theta}}{(1+ \lambda)^{2-a}} \left(1 + \frac{2a-1}{2}
\lambda\right)
\nonumber\\
f_{-2}&=&\frac{2a\lambda \, e^{3i\theta}}{(1+ \lambda)^{3-a}} \left(1 +
2(a-1)\lambda
+ \frac{2a^2-3a+1}{3}\lambda^2\right) \nonumber \end{eqnarray}
The recursion equations for the Laurent coefficients of $\Phi^{(n)}(w)$
can be obtained by
substituting the series of $\Phi$ and $\phi$ into the recursion formula
(\ref{eq-recursion}). We find
\begin{eqnarray}
F^{(n)}_1 &=& F^{(n-1)}_1 f^{(n)}_1
\label{eq-F1-recursion}\\
F^{(n)}_0 &=& F^{(n-1)}_1 f^{(n)}_0 + F^{(n-1)}_0\nonumber\\ F^{(n)}_{-1}
&=& F^{(n-1)}_1 f^{(n)}_{-1} + F^{(n-1)}_{-1}/f^{(n)}_1\nonumber\\ F^{(n)}_{-2}
&=& F^{(n-1)}_1 f^{(n)}_{-2} - F^{(n-1)}_{-1} \frac{f^{(n)}_0}{(f^{(n)}_1)^2}
+ F^{(n-1)}_{-2} \frac{1}{(f^{(n)}_1)^2} \label{eq-coeff-recursion}\\&\cdots&
\nonumber \end{eqnarray}
We note that the $n$-dependence of $f^{(n)}_i$ follows from the dependence
on the randomly chosen $\theta_n$ at the $n$th step, from which follows the
dependence of $\lambda_n$ on $n$. The latter is however a function of all the
previous growth steps, making the iteration
(\ref{eq-F1-recursion}) -(\ref{eq-coeff-recursion}) rather difficult to
analyze.

A general relation between the Laurent coefficients is furnished by the
so-called
area theorem which applies to univalent mappings. Since our maps
solve the Laplace equations with boundary conditions only at infinity and
on the cluster boundary where the potential is zero, they map the $w$ plane
uniquely (and with a unique inverse) to the $z$ plane. In other words, the
pressure lines and the stream lines are non-degenerate. Such mappings have
the property \cite{83Dur} that the area of the image of the unit disc in
the $n$th step is given by: \begin{equation}
S_n =
\left|F^{(n)}_1\right|^2 - \sum_{k=1}^{\infty} k\left|F^{(n)}_{-k}\right|^2
\label{area}
\end{equation}
A second theorem that will be useful in our thinking is a consequence
of the so-called one-fourth theorem, see Appendix A. There
a statement is proven that the interior of the curve
$\{z:z = \Phi^{(n)}(e^{i \theta})\}$ is contained in the
$z$-plane by a circle of radius $4F^{(n)}_1$.  Now as the area
$S_n$ is obtained simply from the superposition of $n$ bumps of roughly the
same area $\lambda_0$, it has to scale like $S_n\approx n \lambda_0$, for large
 $n$. On the other
hand any typical radius of the cluster should scale like
 $n^{1/D}\sqrt{\lambda_0}$ where $D$
is the dimension of the cluster. We can thus
expect a scaling of $F^{(n)}_1$ that goes like
\begin{equation}
F^{(n)}_1\sim n^{1/D}\sqrt{\lambda_0} \ .
\label{eq-scalingrad}
\end{equation}
We note in passing that this scaling law offers us a very convenient
way to measure the fractal dimension of the growing cluster. Indeed,
we measured the dimension D for a range of $a$ in this way by averaging
$F^{(n)}_1$ over 100
clusters. We found that for a range of $a$ spanning the interval $[1/3, 8/9]$
the dimension is constant, around 1.7.

We can infer from Eq.(\ref{eq-scalingrad}) that the sum in
Eq. (\ref{area}) which subtracts positive contributions from $|F^{(n)}_1|^2$
contains terms that cancel the behavior of $n^{2/D}$ (remember
that $D<2$), leaving a power of unity for the scaling of $S_n$.
Indeed, we will show below  both numerical and theoretical evidence for the
 scaling behavior of the
$|F^{(n)}_{-k}|^2$ for $k > 6$ which is in agreement with $n^{2/D}$.

We can give a direct physical interpretation for the coefficients  $F^{(n)}_k$
by comparing them to the coefficients of the series for $\Psi^{(n)}$,
cf. Eq.(\ref{eq-potential}):
\begin{equation}
\Psi^{(n)}(z) = \ln(z) - \ln(r_0) + \sum_1^\infty \frac{\psi_k}{z^k}
\end{equation}
The coefficient of $\ln(z)$ is unity so that the electric flux is unity.
This corresponds to the normalization of the probability.
The constant $r_0$ is the Laplace radius which is the radius of a charged disk
which would give the same field far away. The rest of the $\psi_k$'s are
conventional
multipole moments.

The relations between the Laurent coefficients of  $\Psi^{(n)}$ and
$\Phi^{(n)}$ are:
\begin{eqnarray}
r_0 &=& F_1  \nonumber \\
\psi_{1} &=& -F_0\nonumber \\
\psi_{2} &=& -F_{-1}F_1-\frac{1}{2}F_0^2\nonumber \\
\psi_{3} &=& -F_{-2}F_1^2 -2F_0F_{-1}F_1 -\frac{1}{3}F_{0}^3 \nonumber\\
\psi_{4} &=& -F_{-3}F_{1}^3 -\frac{3}{2} F_{-1}^2F_{1}^2
-3F_{1}F_{0}^2F_{-1}\nonumber\\
&& -3F_{-2}F_{0}F_{1}^2 - \frac{1}{4}F_{0}^4
\end{eqnarray}

The first line shows that $F_1=r_0$, the  Laplace
radius, in accordance with the one-fourth theorem .

The second line shows that the dipole moment $\psi_{1}$ is $-F_0$. We
can interpret this coefficient as a distance, the wandering of the
center of charge due to the random addition of the particles. We will
take the point of view that this quantity is less ``intrinsic'' than the
others to the dynamics of the DLA growth. In fact, if we set
$F_0 =\psi_{1}=0$, (we could imagine shifting the cluster as we grow
it)  we can rewrite the rest of the equations:
\begin{eqnarray}
-F_{-1} & \sim & \psi_{2}/r_0 \nonumber \\
-F_{-2} & \sim & \psi_{3}/r_0^2 \nonumber \\
-F_{-3} & \sim & (\psi_{4}+\frac{3}{2}\psi_{2}^2)/r_0^3 \ ,
\label{Fscale}
\end{eqnarray}
etc. This leads to the interpretation of $F_{-k}$ in terms of the multipole
expansion of the electric field.

\subsection{Statistical objects and the relations to generalized dimensions}

Our growth process is stochastic. Accordingly, it is natural to introduce
averages over the randomness. In our thinking there are two important
averages, one over histories of the whole
random trajectory $\{\theta_i\}_{i=1}^n$,
and the other only over the random choice of $\theta_n$ at the $n$th step.
To distinguish between the two we denote the first by angular brackets and
refer to it as ``history-average'', while the
second is denoted by an overbar and referred to as a ``cluster-average''.
There is a possibility that for very large clusters
($n\to \infty$) the two averages result in the same numbers. We will refer
to such a property as ``self-averaging''.

The cluster average of moments of $\lambda_n$ offers a relationship to the
generalized dimensions of the harmonic measure \cite{83HP}.
The latter are defined by dividing the plane into boxes
of size $\epsilon$, and estimating the probability for a random walker to
hit the piece of the boundary of the cluster which is included in
the $i$th box by
\begin{equation}
p_i(\epsilon) = |E_i| \epsilon \ , \label{pi}
\end{equation}
where $|E_i|$ is the modulus of the electric field $|\nabla \Psi_i|$
at some point in the $i$th  box.
The generalized dimensions are defined by the relation
\begin{equation}
\sum_{i=1}^{N(\epsilon)} p_i^q(\epsilon) \sim \left ({\epsilon\over R}
\right)^{(q-1)D_q}
\end{equation}
where $N(\epsilon)$ is the number of boxes of size $\epsilon$ that are needed
to cover the boundary, and $R$ is the linear size of the largest possible
box, which is of the order of the radius of the cluster.
Substituting (\ref{pi}) we find
\begin{equation}
\epsilon^{q-1}\sum_{i=1}^{N(\epsilon)} |E_i|^q \epsilon\sim
\left ({\epsilon\over R}\right)^{(q-1)D_q}
\end{equation}
Taking $\epsilon$ very small, of the order of $\sqrt{\lambda_0}$, and
assuming that the field is smooth on this scale we have:
\begin{equation}
\int_0^L|E_i|^qds \sim (\sqrt{\lambda_0})^{1-q}~~ n^{(1-q)D_q/D}
\label{gener-dim}
\end{equation}
where $L$ is the length of the boundary, $ds$ is an arc-length differential,
and we have used the scaling law $n\sim S_n/\epsilon^2 \sim (R/\epsilon)^{1/D}$.

The connection to our language is obtained by considering the cluster average
of powers of $\lambda_n$. We grow a cluster of $n-1$ particles,
perform repeated random choices of growth sites (without growing), and
compute  $\lambda_n$
for each choice. The cluster average can be represented as an integral over
the unit circle,
$\overline {\lambda^q_n}$, and is given by
\begin{equation}
\overline {\lambda^q_n} \equiv (1/2\pi)\int_0^{2\pi}\lambda^q_n(\theta) d\theta
 \ .
\end{equation}
Recalling Eq.~(\ref{eq-lambda}) we observe that
$\lambda^q_n (\theta) = \lambda_0^q |E(\theta)|^{2q}$. The last
relation,
Eq.(\ref{dteta}), and Eq.(\ref{gener-dim}) imply the scaling relation
\begin{equation}
\overline {\lambda^q_n} \sim n^{-2qD_{2q+1}/D} \ . \label{lamDq}
\end{equation}

\section{Numerical results and their interpretation}

In this section we present results on three topics:\\
(i) The coefficients of the Laurent expansion. The scaling behavior of
these quantities
is described and discussed in the first subsection.\\
(ii) The microscopic fluctuations in the conformal map.
We show that the assumption of self-averaging is valid
for Eq.~(\ref{lamDq}) and that the multi-fractal exponents are in a good
agreement with the known ones.\\
(iii) Distribution functions of the Laurent coefficients.
We analyze numerically the width of those functions and find
that $F^{(n)}_1$ tends to a deterministic function of $n$. We attribute
this effect
to non-trivial temporal correlations in the field, and give
some evidence of their existence.
\subsection{Laurent Coefficients of $\Phi^{(n)}$}

All the coefficients of the Laurent series of $\Phi^{(n)}(w)$ are complex
numbers except $F_1$ which is real by the choice of zero phase at infinity,
see Eq.(\ref{eq-boundary}).
Most of our discussion below pertains to the amplitudes of the coefficients
$F_k$.
We need to stress, however, that the phases are not irrelevant. If we attempted
to use the correct amplitudes with random phases, the resulting series
will in general not be conformal.

\begin{figure}
\epsfxsize=8.6cm
\centerline{
\epsfbox{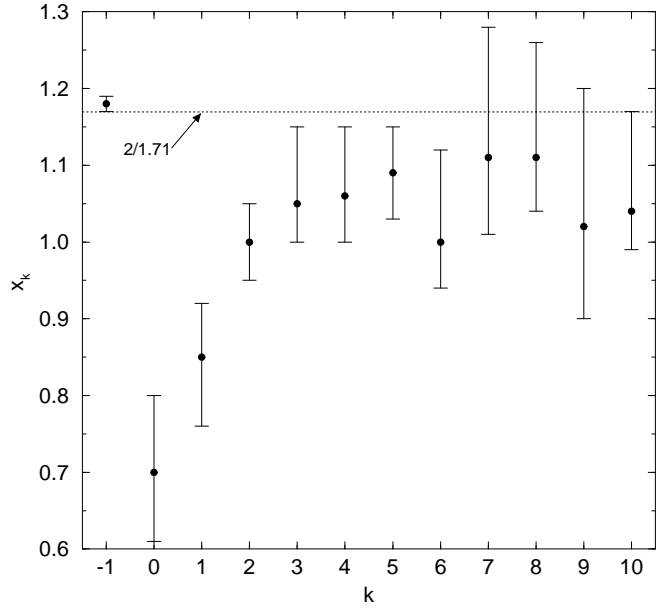}}
\bigskip
\caption{The scaling exponents of the Laurent coefficients:
$\langle|F_{-k}|^2\rangle\sim n^{x_k}$. The values are obtained by averaging
400 independent realizations of 10,000 particle clusters.
}
\label{fig-coeffs}
\end{figure}

One of the main results of this paper is that in a addition to the
expected scaling behavior of the linear coefficient $F^{(n)}_1$
(given in Eq.~(\ref{eq-scalingrad}))
the rest of the amplitudes of the Laurent
coefficients
$|F^{(n)}_{-k}|$ exhibit also a scaling behavior.
We find numerically that in the mean the magnitudes of the Laurent
coefficients scale as
powers of $n$:
\begin{equation}
\langle|F^{(n)}_k|^2\rangle = a_k n^{x_k} \,. \label{scaleFn}
\end{equation}
The exponents $x_k$ are given in Fig.~\ref{fig-coeffs}.
We first discuss the consequences of the scaling behavior of $F^{(n)}_1$.
\subsubsection{Scaling of $F_1$}
\label{sec-sharpness}

The scaling behavior (\ref{eq-scalingrad})
has immediate consequences for the scaling behavior of the bump areas
$\lambda_n$. The connection
appears from the recursion Eq.(\ref{eq-F1-recursion}) of $F_1^{(n)}$
which together with $f_1=(1+\lambda)^a$ reads
\begin{equation}
F_1^{(n)}  = \prod_{k=1}^{n}  [1+\lambda_k]^a  \,. \label{F1a}
\end{equation}
Taking history averages we find
\begin{eqnarray}
\langle F_1^{(n)}\rangle & = &\langle \prod_{k=1}^{n}  [1+\lambda_k]^a\rangle \\
\ln\langle  F^{(n)}_1 \rangle & \approx &  a\sum_{k=1}^{n} \langle
\lambda_k \rangle \\
 d\ln\langle{ F^{(n)}_1}\rangle/dn & \approx & a\langle{\lambda_n}\rangle.
\label{flam}
\end{eqnarray}
The last two equations are obtained by expanding the logarithm and keeping
only divergent sums: both the mean of $F_1^{(n)}$ and the mean of the sum
of $\lambda_k$
increase as a function of $n$. All other sums of powers of $\lambda_k$
converge as a function of $n$, cf. subsection B.
Thus, if we assume that $\langle F^{(n)}_1\rangle  \propto n^{1/D} $,
cf. Eq.(\ref{eq-scalingrad}),
fractal scaling of the radius (see below) implies that \cite{98HL}
\begin {equation}
 \langle{\lambda_n}\rangle = 1/naD \label{meanlam}.
\end{equation}
In the next subsection we show that this is indeed supported by the simulations.
Note that $\langle\lambda_n \rangle$ is inversely proportional to $n$ for any
value of the fractal dimension $D$.
On the other hand, if we assume the property of self-averaging, Eq.
(\ref{meanlam})
implies a
relationship between the generalized dimension
$D_3$ and the fractal dimension $D$. Comparing Eqs.(\ref{lamDq}) and
(\ref{meanlam})
leads immediately to the relation
\begin{equation}
D_3=D/2 \,. \label{D3D}
\end{equation}
This scaling relation was derived by Halsey \cite{87Hal} using much more
elaborate considerations. We see that in the present formalism this
scaling relation is obtained very naturally. In fact the present formulation
is more powerful since Eq.(\ref{meanlam}) predicts not only the
exponent of the third moment of the electric field, but also the prefactor.
It is also noteworthy that the scaling relation (\ref{D3D}) results
simply from the existence of a power law behavior for the radius $F_1^{(n)}$.
\subsubsection{Scaling of $F_0$}

\begin{figure}
\epsfxsize=9.0cm
\centerline{
\epsfbox{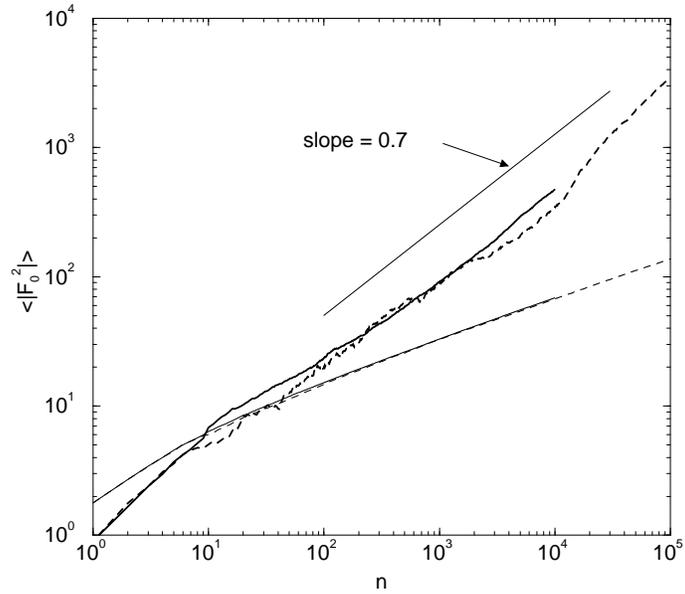}}
\bigskip
\caption{The scaling of $\langle|F_0^{(n)}|^2\rangle$ (thick lines) and the
  sum of diagonal terms (thin lines, see Eq.(51)) with size $n$.
  Clearly the two have different scaling exponents. The solid lines are
averages
over 400 clusters of size 10,000, the dashed lines are averages over 30
clusters
of size 100,000.}
\label{fig-coeff-0}
\end{figure}

We found the exponent of $\langle|F_0|^2\rangle$  to be $x_0 = 0.7\pm0.1$,
see Fig.~\ref{fig-coeff-0}.
To estimate the scaling behavior of $F_0$ theoretically we note that
\begin{equation}
F_0=\frac{1}{2\pi}\int_0^{2\pi}\Phi^{(n)}(\theta)d\theta=
\frac{1}{2\pi}\int_0^L z(s)|E(s)|ds \ .
\end{equation}
Accordingly we can write
\begin{eqnarray}
\label{F0}
\overline{|F_0|^2 } & = & (1/4\pi^2) \int_0^L ds \int_0^L ds' \overline
 {z(s)z(s')^{*}|E(s)||E(s')|} \nonumber \\
 & \sim & \lambda_0 R^2 \int_0^L ds  |E(s)|^2
\end{eqnarray}
In writing the second line we assumed that the main contribution to the
correlation function
is short ranged,
\begin{equation}
\langle z(s)z(s')^{*}|E(s)||E(s')|\rangle \sim \lambda_0 R^2 \overline
 {|E(s)|^2} \delta (s - s') \ . \label{delfunc}
\end{equation}
The justification for this is that the field is expected to exhibit wild
variations as we trace the boundary $z(s)$. In addition the main
contribution to the integral is expected to come from the
support of the harmonic measure where the radius is of the order of $R$.
>From the estimate (\ref{F0}) and Eq.(\ref{gener-dim}) we then find
\begin{equation}
x_0 = \frac{2-D_2}{D} \approx 0.64
\end{equation}
in agreement with our measurement of $x_0$.
(We used here $D_2=0.90$ in correspondence with the numerical finding
reported in Section \ref{ssec-lambda}. Any of
the values of $D_2$ quoted in the literature would yield $x_0$ in the
range $0.7\pm 0.1$.)
\subsubsection{Scaling of $F_{-k}$}

The exponents $x_k$ for  $k <0$
are smaller than $2/D$ but  approach it asymptotically, see
Fig.~\ref{fig-coeffs}.  This behavior is expected from the area
theorem, and also from a direct estimate of the integral
representation of the coefficients for large $k$
\begin{eqnarray}
\label{Fk}
&&\overline {|F_{-k}|^2}  = {1\over 4\pi^2} \int_0^L ds \nonumber\\&&
\times \int_0^L ds' \overline
{ z(s)z(s')^{*}|E(s)||E(s')| e^{i k(\theta (s) - \theta (s')}} \ .
\end{eqnarray}
In Appendix B we show that this integral can be estimated using the multifractal
formalism of the harmonic measure with the final result
\begin{equation}
\overline {|F_{-k}|^2}\sim  (R/4k)^2 \int d\alpha (2k/\pi )^{f(\alpha
)/\alpha}\ ,
\label{Fkalp}
\end{equation}
where $\alpha$ and $f(\alpha)$ are the strength of singularities of the
harmonic measure and the dimension of the sets of points that exhibit these
singularities respectively \cite{86HJKPS}.  For our purposes the important
consequence
of Eq.(\ref{Fkalp}) is the scaling relation (assuming self-averaging)
\begin{equation}
\label{Fkscal}
\langle |F_{-k}|^2 \rangle  = \lambda_0 n^{2/D} g(k)
\end{equation}
with  $g(k) \sim 1/k^2 \int d\alpha  k^{f(\alpha )/\alpha}$. One knows from the
theory of multifractals that $f(\alpha)/\alpha\le 1$, and therefore we can bound
$g(k)$ from above and from below,  $A k^{-2} < g(k) < B k^{-1}$. This is in
accord with our numerical simulations in the range $3\le k\le 10$, although
the calculation in the appendix is only valid for large values of $k$.
We found agreement with Eq.(\ref{scaleFn}) with $x_k\to 2/D$
and $a_k\sim k^{-\alpha}$ with $1<\alpha<2$ .

Note that this scaling behavior has important consequences for both the
area theorem and for conformality. Absolute convergence of the sum
$\sum_{k=1}^{\infty} k|F^{(n)}_{-k}|^2$ in the area theorem requires $\alpha >
2$ which is not the case. The situation is even more serious for the existence
of conformality. To insure the latter the sum $\sum_{k=1}^{\infty}
 k|F^{(n)}_{-k}|$ must exist. This would require $\alpha > 4$. The reason that
the sums exist in the theory is only due to the ultraviolet cutoff
at $\sqrt{\lambda_0}$. This cutoff introduces
a highest $k$ in the Laurent expansion which we estimate as $2\pi
k_{max}\approx
L/\sqrt{\lambda_0} \sim n$ where $L$ is the perimeter of the cluster.

\subsection{Multi-fractal exponents}

\begin{figure}
\epsfxsize=8.6cm
\centerline{
\epsfbox{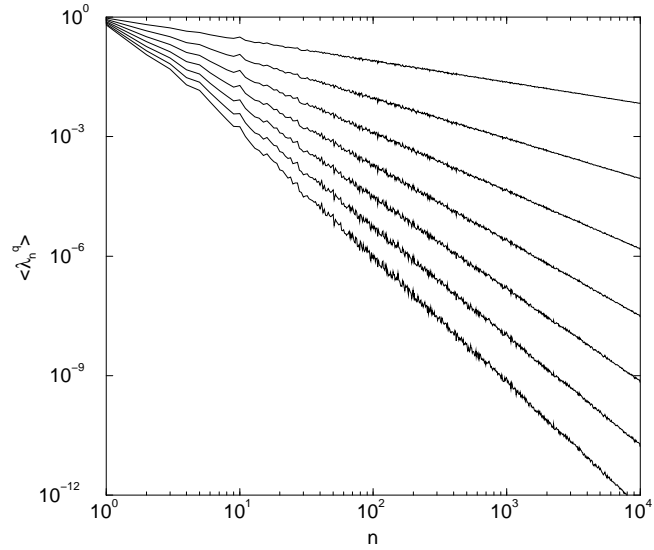}}
\bigskip
\caption{Scaling of the moments $\langle\lambda_n^q\rangle$ with powers of $n$.
  The curves from top to bottom correspond to $q=0.5,\,1,\,1.5,\,2,\,2.5,\,
  3$ and $3.5$.
  The exponents $-2qD_{2q+1}/D$ are in agreement with theoretical predictions
  (see text) and with numerical values for the generalized dimensions in the
  literature.
  }
\label{fig-multifrac}
\end{figure}

Here we test Eq.(\ref{lamDq}).
In Fig.~\ref{fig-multifrac}
we display double-logarithmic plots of $\langle{\lambda^q_n}\rangle $ vs.$n$
for $q=0.5,\,1,\,1.5,\,2,\,2.5,\,3$ and $3.5$.
The values of the exponents derived from our simulations agree very well
(within the uncertainties) with
the generalized dimensions $D_q$ obtained in the past \cite{88Mea}
for $D_2, \cdots, D_8$ using
standard methods. In addition we
reproduce numbers in agreement with the theoretical prediction of
$D_0=D\approx 1.71$
and $D_3=D/2$. This agreement is a strong indication for
self averaging at least for the purpose of computing moments of $\lambda_n$
(i.e. $\langle{\lambda^q_n}\rangle \sim \bar{\lambda^q_n}$).


\subsection{Fluctuations of the averages}
\label{ssec-lambda}

\begin{figure}
\epsfxsize=8.6cm
\centerline{
\epsfbox{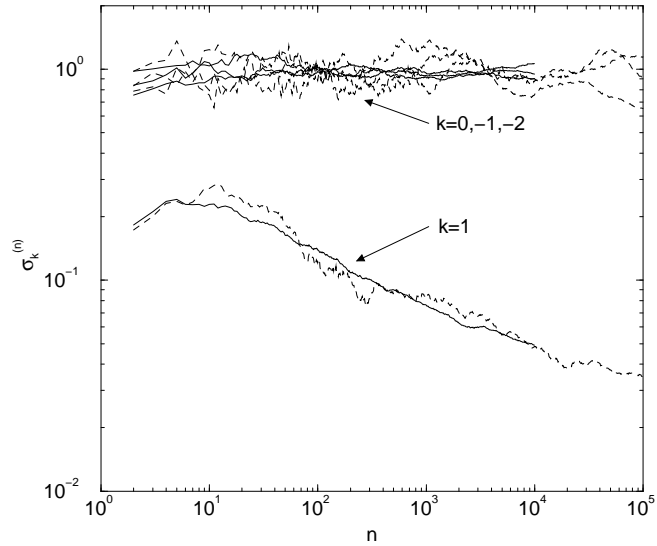}}
\bigskip
\caption{The rescaled standard deviation $\sigma_k^{(n)}$ of the Laurent
  coefficients of the map (see definition in text). For $k\not=1$,
  $\sigma_k^{(n)}$ fluctuates around unity, corresponding to broad
  distributions. For $k=1$ it tends to zero as $n\to\infty$,
  demonstrating the asymptotic sharpness of the distribution of $F_1$.
  The solid lines are averages over 400 clusters of size 10,000, the dashed
  lines are averages over 30 clusters of size 100,000.}
\label{fig-sharpness}
\end{figure}

We previously discussed the scaling behavior of $|F^{(n)}_{-k}|^2$ and showed
that their history averages obey Eq.~(\ref{scaleFn}).
However $|F^{(n)}_{-k}|$ are random
variables with broad scaling distributions.
Fig.~\ref{fig-sharpness} describes the rescaled standard deviation
$\sigma_k^{(n)}$
of the Laurent coefficients,
\begin{equation}
\sigma_k^{(n)}=\sqrt{ \langle {|F_k^{(n)}|}^4 \rangle -
{\langle {|F_k^{(n)}|}^2 \rangle}^2}/\langle {|F_k^{(n)}|}^2 \rangle \ ,
\end{equation}
for $k=1,0,-1,-2$ as a function of the cluster size $n$.
As is seen clearly from the graphs the widths of the
distributions for all $k \le 0$ tend asymptotically to a finite value. This is
the normal behavior for scaling distributions. The exceptional case is $k=1$.
Even though it exhibits a scaling law of the type (\ref{scaleFn})
(see Section III), with
$$x_1 = \frac{2}{D} \approx 1.18 \,,$$
the rescaled distribution width of ${|F_1^{(n)}|}^2$ tends to zero as $n$
goes to infinity. This means that the rescaled distribution function of
$F_1^{(n)}$ tends asymptotically to a delta function. The importance of this
result for the evaluation of the fractal dimension of the cluster warrants
an immediate discussion of this sharpening phenomenon.

The conclusion of the numerics on $F_1$ is that there exists a universal
constant $ c (\lambda_0)$ such that
\begin{equation}
n^{-1/D}F_1^{(n)} \rightarrow c (\lambda_0)
\label{eq-constant}
\end{equation}
where
$c(\lambda_0)$ is cluster independent!
Moreover, we found that
$c(\lambda_0) = c \sqrt{\lambda_0}$,
which is in accordance with the role played by
$\sqrt{\lambda_0}$ as an ultraviolet inner lengthscale, which is
the only lengthscale that appears in the mappings.
Note that the constant $c$ in Eq.(\ref{eq-constant}) depends on the
parameter $a$. We measured $c$ values of 0.6, 0.87, 1.2 and 1.8 for $a$
values of 1/3, 1/2, 2/3 and 4/5 respectively.

The observed sharpening is not obvious since we knowthat $F^{(n)}_1$ is
built from a product of random variables $\lambda_n$, whose moments
change with $n$ in multi-fractal manner according to Eq.~(\ref{lamDq}).

One could attempt to connect the sharpening of $F^{(n)}_1$
to the existence of other sharp functions of $n$.
Considering the full expansion of Eq.(\ref{F1a}) we find
\begin{eqnarray}
\frac{1}{a}\ln {{F_1^{(n)}}} &=& \sum_{i=1}^{n} \ln (1+ \lambda_i) \\
&=& \sum_{i=1}^n \lambda_i - \frac{1}{2} \sum_{i=1}^n \lambda_i^2 +
\frac{1}{3} \sum_{i=1}^n \lambda_i^3 +  \cdots  \ . \label{eq-expand}
\end{eqnarray}
We could understand Eq.~(\ref{eq-constant}) easily if all the sums
of all the powers of $\lambda_i$ converged to constants,
\begin{eqnarray}
\sum_{i=1}^{n} \lambda_i - \frac{2}{D} \ln n &\rightarrow& c_1 \\
\sum_{i=1}^{n} {\lambda_i}^2 &\rightarrow& c_2  \\
\mbox{} &\cdots& \mbox{}
\label{eq-whether}
\end{eqnarray}
with $c_{i}$ cluster independent. In fact, this is not the case.
The sums of powers are not cluster independent.
A clear demonstration of this is a simulation that we performed in
which the initial condition was very far from a circle.  The
individual sums in Eq.~(\ref{eq-expand}) were very different from the average
values, but nevertheless $\sum_{i=1}^{n} \ln (1+ \lambda_i)$
converged to the right value. It is our conclusion that each of the sums
in (\ref{eq-expand}) is not
cluster independent, and yet somehow the resummed form is cluster
independent.

This remarkable sharpening calls for further discussion;
it appears that its interpretation requires better understanding of the time
correlations of
the field: an independent choice of random realization of a series
of $\lambda_i$ according to their multi-fractal distribution can only
generate $F^{(n)}_1$ with the proper scaling exponent but cannot trivially
yield
a highly peaked distribution
of $F^{(n)}_1$. Therefore we consider now some evidence for the
existence of temporal correlations.

The first outstanding evidence appears in the context of the scaling behavior
of $F_0$, which was
discussed in the first subsection. We show that
if we assume that there exist no correlations between different growth
stages, the exponent $x_0$ will be very different from the measured and
calculated value.
>From the recursion relations of the Laurent coefficients
(Eq.~\ref{eq-coeff-recursion})
we can estimate, in the limit of large $n$ when $\lambda_n$ is very small on
the average,
\begin{eqnarray}
  \langle|F^{(n)}_0|^2\rangle &\sim & \sum_{m=1}^n \sum_{m'=1}^n \langle
F_1^{(m)}
  F_1^{(m')}\lambda_m\lambda_{m'}e^{i(\theta_m-\theta_{m'})}\rangle \\
&\sim& \sum_{m=1}^n \sum_{m'=1}^n \langle F_1^{(m)} F_1^{(m')} \rangle
  \langle\lambda_m\lambda_{m'}e^{i(\theta_m-\theta_{m'})}\rangle
\label{eq-F0}
\end{eqnarray}
The second line is obtained because $F_1^{(m)}$ is proportional to the
radius of the whole cluster
and should not be correlated with $\lambda_m$. The crucial approximation comes
next:  if $\lambda_m$ and $\lambda_{m'}$ can
be treated as independent for $m\not=m'$, then (since $\theta_m$ and
$\theta_{m'}$
are independent) Eq.(\ref{eq-F0}) simplifies to
\begin{eqnarray}
&&\langle\lambda_m\lambda_{m'}e^{i(\theta_m-\theta_{m'})}\rangle \approx
  \langle\lambda_m^2\rangle\delta_{m,m'} \label{eq-onlydiag}\\
&&\langle|F^{(n)}_0|^2\rangle \sim \nonumber \\
&&  \sum_{m=1}^n \langle (F_1^{(m))^2}
\rangle \langle\lambda_m^2\rangle
 \sim n^{1+2/D-4D_5/D}\sim n^{0.3}
\label{eq-diag}
\end{eqnarray}
The numerical simulation resulted in an exponent of the order of 0.7,
in serious disagreement with Eq. (\ref{eq-diag}). We think that
the assumption of independence, Eq. (\ref{eq-onlydiag}) is the culprit.

Another fact which illustrates the importance of the time-angle correlation
Eq.~(\ref{eq-onlydiag}) is
the difference between the exponents of $F_0$ and $F_{-1}$
($\langle|F_0|^2\rangle \sim n^{0.7}$ whereas $\langle
|F_{-1}|^2\rangle \sim n^{0.9}$).  Their equations of motion
(\ref{eq-coeff-recursion}) differ, for small $\lambda_n$, by two terms only.
The first one is the term $\lambda_n F^{(n-1)}_{-1}$ in the RHS of
the equation for $F_{-1}$  which is absent in the
equation for $F_0$.
We checked numerically that neglecting this term leads to a very small
change in the exponent. The
second difference between is that the term $\lambda_n \lambda_{n-k}
e^{i (\theta_n - \theta_{n-k})}$ in Eq.(\ref{eq-F0}) is replaced
by $\lambda_n \lambda_{n-k} e^{2
i (\theta_n - \theta_{n-k})}$. The change in the exponent can therefore
be directly attributed to the existence of important time-angle correlations.

\begin{figure}
\epsfxsize=8.6cm
\centerline{
\epsfbox{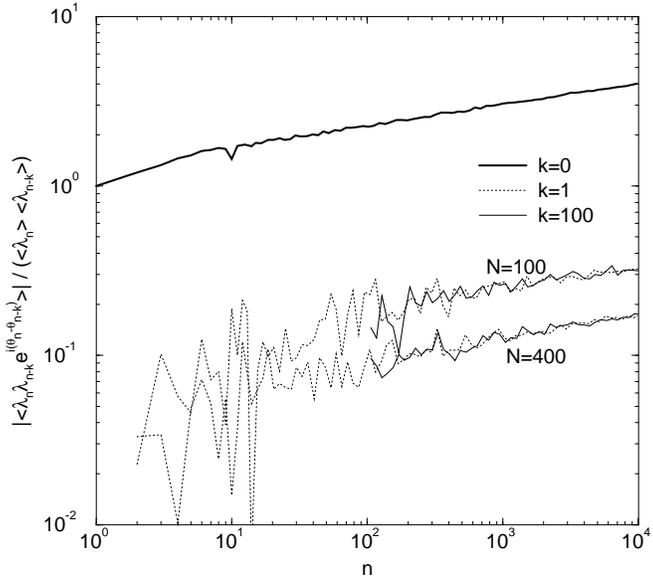}}
\bigskip
\caption{Time-angle correlations of the field.
In order to reduce statistical noise, the
values plotted are averaged in bins $[n, 1.1 n]$.
}
\label{fig-anglecorr}
\end{figure}

We tried to analyze numerically the time-angle correlations
$\langle \lambda_n \lambda_{n-k} e^{i (\theta_n - \theta_{n-k})} \rangle$.
The results for
some $k$'s are shown in Fig.~\ref{fig-anglecorr}. It appears that as we
increase the size of the ensemble, $\langle \lambda_n \lambda_{n-k} e^{i
(\theta_n - \theta_{n-k})} \rangle \to 0$ with the usual $N^{-1/2}$
dependence on the ensemble size. If we believe these numerical results
(doubts may exist due to the relative smallness of the ensembles analyzed),
then the previous results must be related to more subtle correlation of
higher order nature.

\begin{figure}
\epsfxsize=8.6cm
\centerline{
\epsfbox{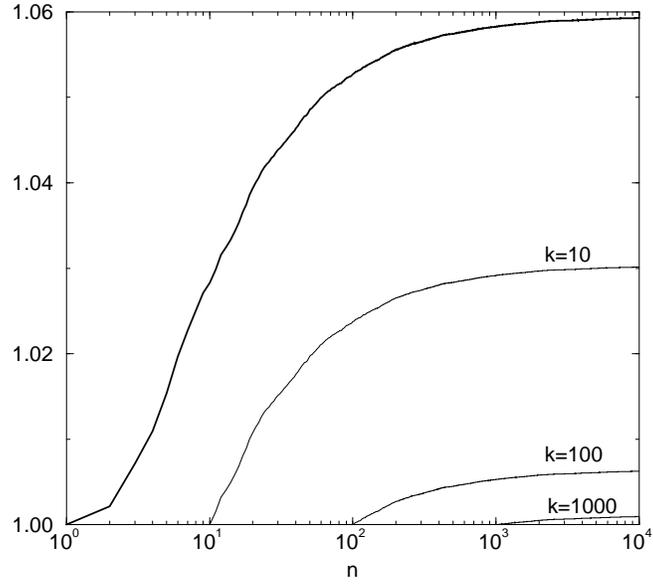}}
\bigskip
\caption{The ratio of $F_1$ approximated by neglecting time correlations and the
full $F_1$:
  $\prod_{i=1}^n \langle(1+\lambda_i)^a\rangle / \langle\prod_{i=1}^n
  (1+\lambda_i)^a\rangle$ (thick line). The quantities
  $\langle\prod_{i=1}^k (1+\lambda_i)^a\rangle \prod_{i=k+1}^n
  \langle (1+\lambda_i)^a\rangle$ are also plotted for $k=10,\,100$ and $1000$.
  }
\label{fig-differ}
\end{figure}

Lastly we would like to discuss the importance of early stages of the
growth. $\langle F_1^{(n)}\rangle$ might be written in the
following way
\begin{equation}
\langle F_1^{(n)}\rangle = \langle\prod_{i=1}^n (1+\lambda_i)^a\rangle \ ,
\end{equation}
(see Eq.~(\ref{eq-F1-recursion})).
Neglecting the correlations in time in the above product one may approximate
\begin{equation}
\langle\prod_{i=1}^n (1+\lambda_i)^a\rangle \approx \prod_{i=1}^n
\langle (1+\lambda_i)^a\rangle  \,.
\label{approx}
\end{equation}
Numerical evaluation of the two objects in Eq.(\ref{approx}) shows that they
differ by a few percent (see Fig.~\ref{fig-differ}). The numerics indicate
the scaling laws
\begin{eqnarray}
\langle\prod_{i=1}^n (1+\lambda_i)^a\rangle &=& c \lambda_{0}
n^{2/D}\ , \label{first}\\
\prod_{i=1}^n \langle(1+\lambda_i)^a\rangle &=& c_1
\lambda_{0} n^{2/D} \ , \label{second}
\end{eqnarray}
where $c_1 / c \geq 1.06$.

To get further intuition we checked also the object
\begin{displaymath}
\langle\prod_{i=1}^k (1+\lambda_i)^a\rangle \prod_{i=k+1}^n
\langle (1+\lambda_i)^a\rangle
\end{displaymath}
for various values of $k$. The results are
shown in Fig.~\ref{fig-differ}. As it seems from this graph,
time correlations between the initial and late stages of the growth
are much more important than local correlations in the late stages.

\begin{figure}
\epsfxsize=8.6cm
\centerline{
\epsfbox{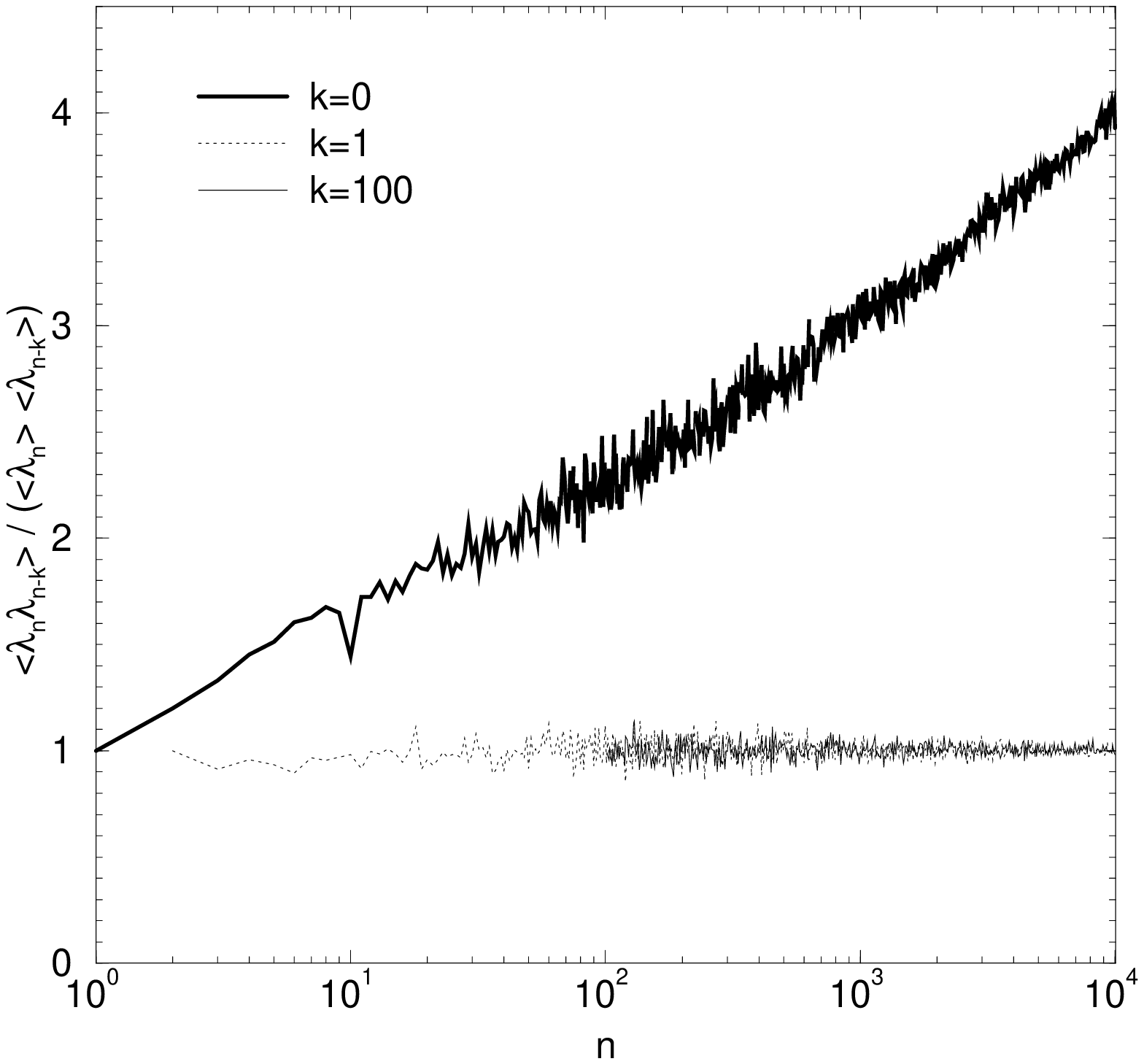}}
\bigskip
\caption{Correlations of the field. In order to reduce statistical noise, the
values plotted are averaged in bins $[n, 1.01 n]$.
}
\label{fig-corr}
\end{figure}

We checked also two-point time correlations $\langle\lambda_n
\lambda_{n-k}\rangle$ for some $k$'s. The results are plotted in
Fig.~\ref{fig-corr}. As it turns out from this graph, $\langle\lambda_n
\lambda_{n-k}\rangle \approx \langle\lambda_n\rangle\langle\lambda_{n-k}
\rangle$ up to statistical fluctuations.


\section{Summary and discussion}

The language proposed by Hastings and Levitov appears to offer many
appealing features. It generates DLA clusters in such a way
that the conformal map $\Phi^{(n)}$ from the circle to the boundary of the
cluster is known at every instant. In this paper we examined
carefully the numerical procedure used to generate the conformal maps,
and pointed out the advantages and the shortcoming of the algorithm.

The new results of this paper pertain to the scaling behavior
of the Laurent coefficients $|F_k|$ of the conformal map $\Phi^{(n)}$
and of the moments of $\lambda_n$ which are related to moments
of the field. We presented a theoretical discussion of the exponents
characterizing moments of $|F_k|$ and  $\lambda_n$. We pointed out the
relations to the multfifractal analysis of the harmonic measure,
and derived scaling relations. Of particular interest is the scaling
relation $D_3=D/2$ that was derived first by Halsey and which appears here
as a very natural consequence of the formalism.

One important result which is not adequately interpreted in this paper is the
sharpness of the distribution of $F_1$. This coefficient is proportional
to the radius of the cluster, and its sharpness is directly related to
the existence of a universal fractal dimension independently of the details
of the shape of the cluster. Understanding the sharpness appears to be
connected to understanding the existence of universal fractal dimension,
and we believe that it poses a very worthwhile and focussed question
for the immediate future.

\acknowledgments
This work has been supported in part by the Israel Science Foundation
founded by the Israel Academy of Sciences and Humanities.
LMS and ES are supported by DOE grant DEFG-02-95ER-45546, and would like to
thank the Weizmann Institute of Science for hospitality.

\appendix
\section{Consequences of the one-fourth theorem}
In this appendix we prove that every univalent function of the type
(\ref{eq-laurent-F}) is bounded in a circle of radius $4 F_1$.
This fact is based on two basic properties of univalent functions \cite{83Dur}:
\begin{enumerate}
\item There is one-to-one correspondence between univalent functions
of the form $f(w) = a_1 w + a_2 w + \cdots$ ($S$-class) and univalent
functions of the form $g(w) = a_1 w + a_{-1}/w + a_{-2}/w^2 + \cdots$
($\Sigma$-class). This correspondence is given by
\begin{equation}
g(w) \leftrightarrow {g(1/w)}^{-1}   \ .
\label{correspondence}
\end{equation}
\item The Koebe One-Quarter Theorem. {\em The image of the unit disc
under every function
of class  S  contains  the disc} ${z:|z| < 4 |a_1|}$.
\end{enumerate}
Consider a function $\Phi (w)$ of the form (\ref{eq-laurent-F}).
This is a $\Sigma$-class function with linear coefficient $F_1$.
Let us denote its conjugate (By Eq.~(\ref{correspondence})) $S$-class function
as $P(w)$. The linear coefficient of $P$ is $1/F_1$. Consider now the
smallest circle in the $z-$plane which bounds the image of the unit circle
under $\Phi$, $\{z: |z| = R\}$. From Eq.~\ref{correspondence} it is clear that
the circle $\{z: |z| = 1/R\}$ is the largest circle which is contained in the
image of the unit disc under $P$. Thus the Koebe One-Quarter Theorem ensures
that $1/R \geq 1/(4F_1)$,
which implies $R \leq 4F_1$.
\section{Estimate of the scaling behavior of $\langle |
 F_{-k}^{(n)}|^2 \rangle$}

To estimate the large $k$ and large $n$ dependence of $F^{(n)}_{-k}$,
the components are first written as integrals over
the boundary of the cluster
\begin{eqnarray}
\label{A1}
F^{(n)}_{-k} & = & (1/2\pi )\int_0^{2\pi}\Phi^{(n)}(e^{i\theta}) e^{i k
\theta} d\theta
 \nonumber \\
& = & \int_0^{L} z(s) e^{i k \theta (s)} |E(s)| ds
\end{eqnarray}
where
\begin{equation}
\label{A2}
\theta (s) = \int_0^s |E(s')| ds'.
\end{equation}
For the purposes of Sect. 4 we are interested in $|F^{(n)}_k|^2$:
\begin{equation}
|F^{(n)}_k|^2=\int_0^{L}\!\!\int_0^{L}\!\! z(s) z*(s') e^{i k [\theta (s)-
\theta (s')]} |E(s)|
|E(s')| dsds' \ . \label{estim}
\end{equation}
For a given value of $n$ (or equivalently, of $R\sim n^{1/D}$),
an examination of Eq (\ref{estim}) shows that for large $k$ the
fluctuations in the values of the integrands depend more crucially on
the phase variations than on the field and radius variations. The phase varies
appreciably when $\theta$ changes an amount
\begin{equation}
\label{A3}
\Delta \theta \approx (\pi /2k).
\end{equation}
and therefore it is useful to split up the integral Eq.(\ref{A1}) into a sum of
 essentially independent contributions coming
from the electric field singularities with exponents $\alpha$. This exponent
is determined by the scaling law relating the measure (which is
proportional to $\Delta \theta$) of a box to its size
$(\Delta s)_\alpha$: $\Delta \theta \sim ((\Delta s)_\alpha/R)^\alpha$
\cite{86HJKPS}.
The integral is split into
 contributions made of contour sections of different
lengths $(\Delta s)_{\alpha} $ dependent on the singularity but each giving
 rise  to the same change $\Delta \theta$.
If one can estimate both  the magnitude of the contribution of
a specific  multifractal electric field singularity $\alpha$ to the integral
and the number of such
contributions $ {\cal N}_{\alpha}(k,n)$ \cite{86HJKPS}, then one can write
\begin{eqnarray}
\label{A4}
\langle |F_{-k}^{(n)}|^2\rangle & \sim & \sum_j |I_{\alpha_j}(k,n)|^2
\nonumber \\
& \sim & \int d\alpha {\cal N}_{\alpha}(k,n) |I_{\alpha}(k,n)|^2 \ ,
\end{eqnarray}
where
\begin{equation}
I_\alpha(k,n)\equiv \int_{(\Delta S)_\alpha} |E(s)|z(s) ds \ . \label{Ialph}
\end{equation}
To  estimate  ${\cal N}_{\alpha}(k,n)$ we recall that by definition
\begin{equation}
\label{A5}
\Delta \theta = \int_s^{s +\Delta s} |E(s')|ds' \sim (\Delta s /R)^{\alpha} \ .
\end{equation}
>From Eq.(\ref{A3})
\begin{equation}
\label{A6}
(\Delta s)_{\alpha} \sim R (\pi/2 k)^{1/\alpha} .
\end{equation}
Using Eq.(\ref{A6}) and the fact that the fractal dimension of
singularities of
 size $\alpha$ is $f(\alpha )$,
we can now also estimate the number of singularities of size $\alpha$ which
 contribute to the integral as
\begin{equation}
\label{A7}
{\cal N}_{\alpha}(k) \sim (R/(\Delta s)_{\alpha})^{f(\alpha )} \sim (2k/\pi
 )^{f(\alpha )/\alpha}
\end{equation}
To estimate  $|I_{\alpha}(k,n)|^2$ we note that the major contribution to
Eq. (\ref{Ialph}) comes from the support of the harmonic measure where
$|z(s)|\approx R$. Accordingly
\begin{equation}
\label{A9}
|I_{\alpha}|^2 \sim R^2\left[\int_{(\Delta s)_\alpha} |E(s)| ds\right]^2\sim
R^2 (\Delta \theta)^2\sim R^2 (2\pi/k)^2 \ ,
\end{equation}
where we have made use of Eqs.(\ref{A2}), (\ref{A3}).
Combining Eq.(\ref{A4}) with the estimates (\ref{A7}) and (\ref{A9}) then yields
\begin{equation}
\label{A11}
\langle |F_{-k}^{(n)}|^2\rangle \sim  (R/4k)^2\int d\alpha (2k/\pi )^{f(\alpha
 )/\alpha} \ .
\end{equation}
We note that the approximation adopted in this appendix differs from the
delta-function assumption (\ref{delfunc}) in asserting that for high
values of $k$ the variation of the phase dominates the decay of the
integrand compared to the rapid decorrelation of the field. One would guess
that for $k$ of the order of unity the field decorrelates faster due to the
rapid variation over the arc length. For high values of $k$ the phase
decorrelation
is strongly amplified and we adopt the assumption used here.


\end{multicols}

\end{document}